\documentclass[twocolumn,floatfix,showpacs,showkeys,preprintnumbers,nofootinbib,superscriptaddress]{revtex4}
\usepackage[utf8]{inputenc}
\usepackage[sort&compress]{natbib}
\usepackage{ulem}
\usepackage{bm}
\usepackage{times}
\usepackage{amssymb,amsbsy,amsmath,amsfonts}
\usepackage{graphicx}
\usepackage{float}
\usepackage{color}
\usepackage{morefloats}
\usepackage{rotating}
\usepackage{srcltx}
\usepackage{overpic}
\usepackage{slashed}
\usepackage{subfigure}
\usepackage{multirow}
\usepackage{verbatim}
\usepackage{hyperref}
\usepackage{tabularx}

\begin{document}

\title{Three-body molecules $\bar{D}\bar{D}^{\ast}\Sigma_{c}$- understanding the nature of $T_{cc}$, $P_{c}(4312)$, $P_{c}(4440)$ and $P_{c}(4457)$ }

\author{Ya-Wen Pan}
\affiliation{School of Physics, Beihang University, Beijing 102206, China}

\author{Tian-Wei Wu}
\affiliation{School of Fundamental Physics and Mathematical Sciences, Hangzhou Institute for Advanced Study, UCAS, Hangzhou, 310024, China}

\author{Ming-Zhu Liu}\email{ zhengmz11@buaa.edu.cn}
\affiliation{School of Space and Environment, Beihang University, Beijing 102206, China}
\affiliation{School of Physics, Beihang University, Beijing 102206, China}

\author{Li-Sheng Geng}\email{ lisheng.geng@buaa.edu.cn}
\affiliation{School of Physics, Beihang University, Beijing 102206, China}
\affiliation{
Beijing Key Laboratory of Advanced Nuclear Materials and Physics,
Beihang University, Beijing 102206, China}
\affiliation{School of Physics and Microelectronics, Zhengzhou University, Zhengzhou, Henan 450001, China}

\begin{abstract}
The nature of the three pentaquark states, $P_{c}(4312)$,  $P_{c}(4440)$ and $P_{c}(4457)$, discovered by the LHCb Collaboration in 2019, is still under debate, although the $\bar{D}^{(\ast)}\Sigma_{c}$ molecular interpretation seems to be the most popular. In this work, by adding a $\bar{D}$ meson into  the $\bar{D}^{\ast}\Sigma_{c}$ pair, we  investigate the mass and decay width of the three-body  molecules $\bar{D}\bar{D}^{\ast}\Sigma_{c}$ and explore  the correlation between the existence  of the $\bar{D}\bar{D}^{\ast}\Sigma_{c}$ molecules with the existence of $\bar{D}^{(\ast)}\Sigma_{c}$ and $\bar{D}^{\ast}\bar{D}$ two-body molecules. The latter  can be identified  with the doubly charmed tetraquark state $T_{cc}$ recently discovered by the LHCb Collaboration. Based on the molecular nature of $P_{c}(4312)$,  $P_{c}(4440)$,  $P_{c}(4457)$, and $T_{cc}$, 
our results indicate that there exist two three-body bound states of $\bar{D}\bar{D}^{\ast}\Sigma_{c}$ with $I(J^{P})=1(1/2^{+})$ and $I(J^{P})=1(3/2^{+})$,  and  binding energies $37.24$ MeV and $29.63$ MeV below the $\bar{D}\bar{D}^{\ast}\Sigma_{c}$ mass threshold. In addition, we find that the mass splitting of these two three-body molecules are correlated to the mass splitting of $P_{c}(4440)$ and $P_{c}(4457)$, which 
offers a non-trivial way to reveal the molecular nature of these states.  The partial  widths of two $\bar{D}\bar{D}^{\ast}\Sigma_{c}$ molecules  decaying into $J/\psi p \bar{D}$ and $J/\psi p \bar{D}^{\ast}$ are found to be several MeV.  We recommend  the experimental searches for the $\bar{D}\bar{D}^{\ast}\Sigma_{c}$ molecules in the $J/\psi p \bar{D}$ and $J/\psi p \bar{D}^{\ast}$ invariant mass distributions.

\end{abstract}
\maketitle

\section{Introduction}
   In terms of the constituent quark model proposed by Gell-mann~\cite{Gell-Mann:1964ewy} and Zweig~\cite{Zweig:1964ruk,Zweig:1964jf}, hadrons can be classified either as mesons made of a pair of quark and anti-quark or baryons made of three quarks, the property of which can be well described in the conventional quark model~\cite{Godfrey:1985xj,Capstick:1985xss}.   However, more and more the so-called exotic states beyond the traditional quark model have been discovered  experimentally,  starting from $X(3872)$  in 2003~\cite{Choi:2003ue}. To clarify the  nature of these exotic states, a lot of theoretical interpretations were proposed,     such as hadronic molecules, compact multiplet quark sates, kinetic effects, and so on (for recent reviews, see Refs.~\cite{Chen:2016qju,Hosaka:2016ypm,Lebed:2016hpi,Oset:2016lyh,Guo:2017jvc,Olsen:2017bmm,Ali:2017jda,Brambilla:2019esw,Liu:2019zoy,Guo:2019twa}  ). Among them, the hadronic molecular picture is rather popular because many (if not all) of these states are located near the mass threshold of a pair of conventional hadrons. Nevertheless, how to confirm the molecular  nature of these exotic states remains a big challenge for both experiments and theory.  

 To confirm an exotic state as  a hadronic molecule, one needs to be able to describe its production rate, decay width,  mass, spin-parity, and other relevant properties consistently in the molecular picture.   However,  most approaches can  only describe part of the relevant properties. This motivates us to find alternative methods to help achieve this goal.   In nuclear physics,  the existence of light nuclei, such as triton or $^3H$  serves as a non-trivial check on the two-body bound-state nature of the deuteron. 
Along this line,  assuming  $D_{s0}^*(2317)$ as a $DK$ bound state, we have studied the few-body systems of $DDK$ and $DDDK$, the existence of which indeed support the molecular nature of $D_{s0}^*(2317)$~\cite{Wu:2019vsy}. 
 In this work, assuming   $T_{cc}$ and the three pentaquark states [$P_{c}(4312)$, $P_{c}(4440)$ and  $P_{c}(4457)$] as $DD^{\ast}$ and $\bar{D}^{\ast}\Sigma_{c}$ bound states, respectively,  we  investigate the related three-body system $\bar{D}\bar{D}^{\ast}\Sigma_{c}$  and explore the correlation of  the three-body bound states $\bar{D}\bar{D}^{\ast}\Sigma_{c}$ with the related two-body bound states.  

The hidden charm pentaquark states, $P_{c}(4380)$ and $P_{c}(4450)$,  were firstly discovered by the LHCb Collaboration in 2015~\cite{Aaij:2015tga}. With a statistics  10 times larger, the  $P_{c}(4450)$ state splits into $P_{c}(4440)$ and $P_{c}(4457)$, in addition a new state $P_{c}(4312)$ appears, all of which lie close to the mass thresholds of $\bar{D}^{(\ast)}\Sigma_{c}$~\cite{Aaij:2019vzc}. In our previous work~\cite{Liu:2019tjn}, we have employed a contact-range  effective field theory (EFT) to assign  $P_{c}(4312)$,  $P_{c}(4440)$ and $P_{c}(4457)$ as $\bar{D}^{(\ast)}\Sigma_{c}$ hadronic molecules dictated by heavy quark spin symmetry(HQSS), which was confirmed by many other groups~\cite{Xiao:2019aya,Xiao:2019mvs,Sakai:2019qph,Yamaguchi:2019seo,Liu:2019zvb,Valderrama:2019chc,Meng:2019ilv,Du:2019pij,Burns:2019iih,Wu:2019rog,Azizi:2020ogm,Phumphan:2021tta}. Even so, there still exist other explanations, such as, hadro-charmonium~\cite{Eides:2019tgv}, compact pentaquark states~\cite{Ali:2019npk,Mutuk:2019snd,Wang:2019got,Cheng:2019obk,Weng:2019ynv,Zhu:2019iwm,Pimikov:2019dyr,Ruangyoo:2021aoi}, virtual states~\cite{Fernandez-Ramirez:2019koa} and double triangle singularities~\cite{Nakamura:2021qvy}. The existence of three-body bound states $\bar{D}\bar{D}^{\ast}\Sigma_{c}$ will further verify the molecular nature of the pentaquark states, where $\bar{D}\bar{D}^{\ast}\Sigma_{c}$ system can be viewed as a cluster of a $\bar{D} $ meson and the $\bar{D}^{\ast}\Sigma_{c}$ pair or  a $\bar{D}^{\ast} $ meson and the $\bar{D}\Sigma_{c}$ pair.

In addition, the $\bar{D}\bar{D}^{\ast}\Sigma_{c}$ system can be regarded as a cluster of a $\Sigma_{c}$ baryon and the $\bar{D}^{\ast}\bar{D}$ pair, which is related to the doubly charmed tetraquark state ${T}^+_{cc}$ discovered by the LHCb Collaboration ~\cite{LHCb:2021vvq}.
  The mass of $T^+_{cc}$ is below the mass threshold of $D^{0}D^{\ast+}$ by only several hundred keV, and its decay width from the unitary analysis  is rather small, only a few tens of keV~\cite{LHCb:2021auc}.  Assuming that the $T_{cc}$ state is a $DD^{\ast}$ bound state,  its mass and decay width can be described in the hadronic molecular model~\cite{Meng:2021jnw,Dong:2021bvy,Chen:2021vhg,Ling:2021bir,Ren:2021dsi,Feijoo:2021ppq,Yan:2021wdl,Albaladejo:2021vln,Du:2021zzh,Ke:2021rxd}. Although the molecular interpretation seems to be the most popular, the interpretation of a compact tetraquark state cannot be ruled out.  The study of the three-body system $\bar{D}\bar{D}^{\ast}\Sigma_{c}$ could also be helpful to verify its molecular nature.

The three-body system  $\bar{D}\bar{D}^{(\ast)}\Sigma_{c}$ is particularly interesting for a number of reasons. First of all, there is no annihilation of a pair of light quark and its anti-quark. This indicates that such a state, if exists,   has a minimum quark content $\bar{c}\bar{c}cqqqq$, which is explicitly exotic. Second, the interactions of the sub-systems  $\bar{D}\Sigma_{c}$,  $\bar{D}^{\ast}\Sigma_{c}$, and $\bar{D}^{\ast}\bar{D}$ can be precisely 
determined by reproducing the masses of their corresponding molecular candidates using the one-boson-exchange (OBE) potential, which largely reduces  the uncertainty of  the so-obtained binding energy of the $\bar{D}\bar{D}^{\ast}\Sigma_{c}$ state.  Such exotic states, if discovered by experiments, will help verify the molecular nature of $T_{cc}$ as well as $P_{c}(4312)$,  $P_{c}(4440)$ and $P_{c}(4457)$. In this work, we employ the Gaussian Expansion Method(GEM) to study the three-body system, $\bar{D}\bar{D}^{\ast}\Sigma_{c}$, and then use an effective Lagrangian approach to evaluate its main strong decay modes. 

This paper is organized as follows.  In Sec.~\ref{sec:formalism}, we briefly explain how to solve the three-body Schr\"{o}dinger equation by GEM. Next in Sec.~\ref{sec:pre}, we present the binding energies of the three-body bound states $\bar{D}\bar{D}^{\ast}\Sigma_{c}$ and calculate the partial widths of  the $\bar{D}\bar{D}^{\ast}\Sigma_{c}$ molecules decaying into $J/\psi p\bar{D}^{(\ast)}$ and $\bar{T}_{cc} \Lambda_{c}\pi$.   Finally, this paper is ended with a short summary in Sec.~\ref{sum}.

\section{FORMALISM}
\label{sec:formalism}

\begin{figure}[ttt]
\begin{center}
\begin{overpic}[scale=.153]{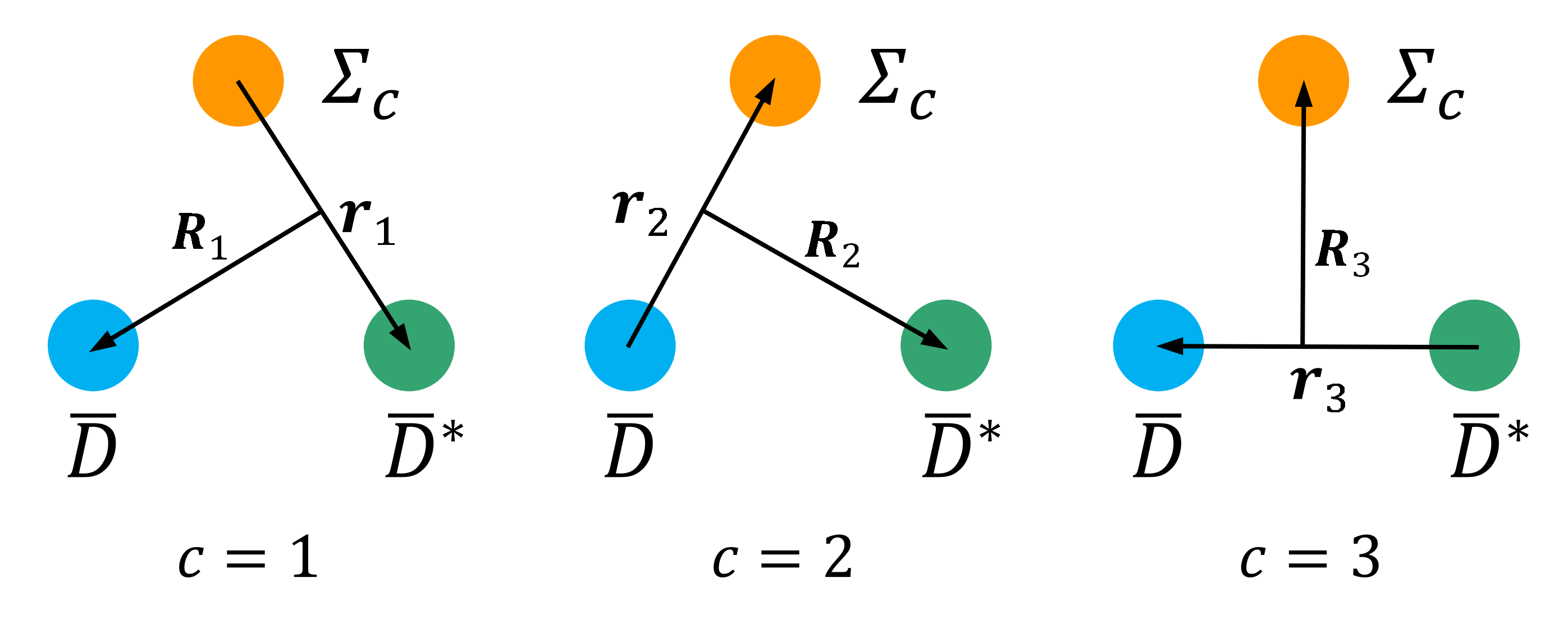}
\end{overpic}
\caption{ Three Jacobi coordinates of the $\bar{D}\bar{D}^{*}\Sigma_{c}$ system }
\label{DDSigmaC}
\end{center}
\end{figure}

To obtain the binding energy of the $\bar{D}\bar{D}^{(\ast)}\Sigma_{c}$ system, we need to solve the three-body Schr\"{o}dinger equation by GEM, which has been widely applied to investigate few-body systems in nuclear physics~\cite{Hiyama:2003cu} and hadron physics~\cite{Hiyama:2005cf,Yoshida:2015tia}. The three-body Schr\"{o}dinger equation reads
\begin{equation}
    [T+V^{1}(r_{1})+V^{2}(r_{2})+V^{3}(r_{3})-E]\Psi_{JM}^{Total}=0,
\end{equation}
where $T$ is the kinetic-energy operator, $V^{i}(r_{i})$ is the potential between the $j_{\rm{th}}$ and $k_{\rm{th}}$ particle pair $(i,j,k=1-3)$,  and the $1_{\rm{st}}$, $2_{\rm{nd}}$ and $3_{\rm{rd}}$ particle refer to the $\bar{D}$ meson, $\bar{D}^{*}$ meson and $\Sigma_{c}$ baryon, respectively.   
The total wave function $\Psi_{JM}^{Total}$  is expressed as a sum of three component functions:
\begin{equation}
    \Psi_{JM}^{Total}=\sum_{i=1}^{3}C_{i,\alpha}\Phi_{JM,\alpha}^{c=i}(\textbf{r}_{i},\textbf{R}_{i}),
\end{equation}
where $C_{i,\alpha}$    are the expansion coefficients of relevant basis, $i=1,2,3$ denotes the three channels of Fig.~\ref{DDSigmaC}, and $\alpha\equiv\{nl,NL,\lambda,\Sigma,s,T,t\}$. Here $l$ and $L$ are the orbital angular momentum of the coordinates $r$ and $R$, $t$ and $s$ are the isospin and spin of the two-body subsystem in each channel, and $\lambda$, $\Sigma$ and $T$ are the total orbital angular momentum, spin and isospin, respectively. The wave function of each channel is expressed as 
\begin{equation}
    \Phi_{JM,\alpha}^{c}(\textbf{r}_{i},\textbf{R}_{i})=\left[\Phi_{lL,\lambda}^{c}\Omega_{\Sigma,s}^{c}\right]_{JM}H_{Tt}^{c},
\end{equation}
where $\Phi_{lL,\lambda}^{c}$ is the spatial wave function, and $\Omega_{\Sigma,s}^{c}$ is the spin wave function.  The total isospin wave function $H_{Tt}^{c} $ in each channel are written as 
\begin{equation}
   \begin{aligned}
    &H_{Tt_{1}}^{c=1}=[[\eta_{\frac{1}{2}}(\bar{D}^{\ast})\eta_{1}(\Sigma_{c})]_{t_{1}}\eta_{\frac{1}{2}}(\bar{D})]_{T},\\
    &H_{Tt_{2}}^{c=2}=[[\eta_{\frac{1}{2}}(\bar{D})\eta_{1}(\Sigma_{c})]_{t_{2}}\eta_{\frac{1}{2}}(\bar{D}^{\ast})]_{T},\\
    &H_{Tt_{3}}^{c=3}=[[\eta_{\frac{1}{2}}(\bar{D}^{\ast})\eta_{\frac{1}{2}}(\bar{D})]_{t_{3}}\eta_{1}(\Sigma_{c})]_{T},
   \end{aligned}
\end{equation}
where $\eta$ is the isospin wave function of each particle. The spatial wave function $\Phi_{lL,\lambda}^{c}$ can be expanded as
\begin{equation}
   \begin{aligned}
    &\Phi_{lL,\lambda}^{c}=[\phi_{n_{c}l_{c}}^{G}(\textbf{r}_{c})\psi_{N_{c}L_{c}}^{G}(\textbf{R}_{c})]_{\lambda},\\
    &\phi_{nlm}^{G}(\textbf{r})=N_{nl}r^{l}e^{-\nu_{n}r^{2}}Y_{lm}(\hat{\textbf{r}}),\\
    &\psi_{NLM}^{G}(\textbf{R})=N_{NL}R^{L}e^{-\lambda_{N}R^{2}}Y_{LM}(\hat{\textbf{R}}),
   \end{aligned}
\end{equation}
where  $N_{nl}(N_{NL})$  is the  normalization constant, and  the relevant parameters $\nu_{n}$ and $\lambda_{N}$ are given by
\begin{equation}
   \begin{aligned}
    &\nu_{n}=1/r^{2}_{n},\quad r_{n}=r_{1}a^{n-1},\quad (n=1-n_{max}),\\
    &\lambda_{N}=1/R^{2}_{N},\quad R_{N}=R_{1}A^{N-1},\quad (N=1-N_{max}),
   \end{aligned}
\end{equation}
where $\{n_{max}, r_{min}, a~ \mbox{or}~$$r_{max} \}$ and $\{N_{max}, R_{min},  A~ \mbox{or}~ $ $ R_{max} \}$ are Gaussian basis parameters given in Table~\ref{quantum number}.

\begin{table}[ttt]
    \centering
    \caption{Three-body angular-momentum space and the Gaussian range parameters for the $I(J^{P}) = 1(\frac{1}{2}^{+})$ and $1(\frac{3}{2}^{+})$ configurations of the $\bar{D}\bar{D}^{*}\Sigma_{c}$ system. Lengths are in units of fm.\label{quantum number}}
    \begin{tabular}{ccccccccccc}
    \hline\hline
     \multirow{2}{*}{$I(J^{P})$}& \multirow{2}{*}{$c$} & \multirow{2}{*}{$l$} & \multirow{2}{*}{$L$} & \multirow{2}{*}{$\lambda$} & \multirow{2}{*}{$s$} & \multirow{2}{*}{$\Sigma$} & \multirow{2}{*}{$t$} &$n_{max}$ & $r_{min}$ & $r_{max}$  \\
     &  &  &  &  &  &  &  &$(N_{max})$ & $(R_{min})$ & $(R_{max})$  \\\hline
     \multirow{3}{*}{$1(\frac{1}{2}^{+})$}&1 & 0 & 0 & 0 & 1/2 & 1/2 & 1/2(3/2) & 10 & 0.1 & 20.0  \\
     ~&2 & 0 & 0 & 0 & 1/2 & 1/2 & 1/2(3/2) & 10 & 0.1 & 20.0 \\
     ~&3 & 0 & 0 & 0 & 1 & 1/2 & 0(1) & 10 & 0.1 & 20.0 \\\hline
    \multirow{3}{*}{$1(\frac{3}{2}^{+})$}&1 & 0 & 0 & 0 & 3/2 & 3/2 & 1/2(3/2) & 10 & 0.1 & 20.0  \\
      ~&2 & 0 & 0 & 0 & 1/2 & 3/2 & 1/2(3/2) & 10 & 0.1 & 20.0 \\
     ~&3 & 0 & 0 & 0 & 1 & 3/2 & 0(1) & 10 & 0.1 & 20.0  \\
    \hline\hline
    \end{tabular}\\
\end{table}

\section{Results and Discussions}
\label{sec:pre}

First we discuss the  quantum numbers of the $\bar{D}\bar{D}^{\ast}\Sigma_{c}$ system.   
Considering only $S$-wave interactions,  the total angular momentum of the $\bar{D}\bar{D}^{\ast}\Sigma_{c}$ system is either $J=1/2$ or $J=3/2$.    The isospin of $\bar{D}\bar{D}^{\ast}$ is either 0 or 1. In the OBE model, the interaction in isospin 0 is much stronger than that in isospin 1, to such an extent that $T_{cc}$ can be understood as an isospin 0 $\bar{D}\bar{D}^{\ast}$ bound state. As a result, the total isospin of the $\bar{D}\bar{D}^{\ast}\Sigma_{c}$ system is taken to be 1. Therefore, in this work we investigate the two  $\bar{D}\bar{D}^{\ast}\Sigma_{c}$ configurations  with $I(J^{P})=1(\frac{1}{2}^{+})$ and $I(J^{P})=1(\frac{3}{2}^{+})$. The relevant quantum numbers of  the $\bar{D}\bar{D}^{\ast}\Sigma_{c}$ system are given Table~\ref{quantum number}.

 In this work,  we employ the OBE model to construct  the potentials of $\bar{D}\Sigma_{c}$,  $\bar{D}^{\ast}\Sigma_{c}$, and  $\bar{D}^{\ast}\bar{D}$ through the effective Lagrangians describing the interactions between charmed hadrons and light mesons $\pi$, $\rho$, $\sigma$ and $\omega$.  For details, we refer to Refs.~\cite{Liu:2019stu,Liu:2019zvb}. Since the $S$-wave interaction plays a dominant  role in forming hadronic molecules, we only consider the $S$-wave interaction in this work.   To estimate the impact of the finite size of hadrons on the OBE potentials, we adopt  a monopole form factor $\frac{\Lambda^2-m^2}{\Lambda^2-q^2}$ for the relevant meson-baryon vertices, which introduces an unknown parameter $\Lambda$. To decrease the uncertainty of the OBE potential induced by the cutoff, we determine it by reproducing the masses of some well known molecular candidates.  
Assuming  $P_{c}(4312)$, $P_{c}(4440)$, and $P_{c}(4457)$ as $\bar{D}^{(\ast)}\Sigma_{c}$ bound states, the corresponding cutoff (denoted by $\Lambda_{P}$) is fixed  to  be 1.16 GeV, while the cutoff of the $\bar{D}\bar{D}^{\ast}$ system (denoted by $\Lambda_{T}$)  is fixed
to be $0.998$ GeV  if $T_{cc}$ is regarded  as a $\bar{D}\bar{D}^{\ast}$ bound state. Therefore, we  take three sets of cutoff values to search for three-body bound states in the  $\bar{D}\bar{D}^{\ast}\Sigma_{c}$ system:
Case I: $\Lambda_{T}$=$\Lambda_{P}$=0.998 GeV; Case II: $\Lambda_{T}$=$0.998$ GeV,~$\Lambda_{P}$=1.16 GeV; and Case III: $\Lambda_{T}$=$\Lambda_{P}$=1.16 GeV. As the OBE interaction increases with the cutoff, we anticipate that Case III will yield the largest binding energies while Case I the smallest ones.


\begin{table}[ttt]
    \centering
    \caption{Binding energies (in units of MeV), expectation values of the Hamiltonian (potential and kinetic energies)(in units of MeV) and root-mean-square radii (in units of fm) of the three-body system $\bar{D}\bar{D}^{*}\Sigma_{c}$ obtained in the three cases detailed in the main text. \label{results1} }
    \begin{tabular}{ccccccccc}
    \hline\hline
     $I(J^{P})$& $B$ & $T$ & $V_{\bar{D}^{*}\Sigma_{c}}$ & $V_{\bar{D}\Sigma_{c}}$ & $V_{\bar{D}\bar{D}^{*}}$ & $r_{\bar{D}^{*}\Sigma_{c}}$ & $r_{\bar{D}\Sigma_{c}}$ & $r_{\bar{D}\bar{D}^{*}}$   \\\hline\hline
     \multicolumn{9}{c}{Case I ~~~~~~$\Lambda_{P}=\Lambda_{T}=0.998$ GeV}\\\hline
     $1(\frac{1}{2}^{+})$ & 10.86 & 65.41 & -19.64 & -21.69 & -34.94 & 1.42 & 1.41 & 1.36 \\\hline
     $1(\frac{3}{2}^{+})$ & 7.06 & 52.18 & -19.66 & -10.46 & -29.12 & 1.62 & 1.81 & 1.64 \\\hline\hline
     \multicolumn{9}{c}{Case II ~~~~~~$\Lambda_{T}=0.998$ GeV\quad$\Lambda_{P}=1.16$  GeV}\\\hline
     $1(\frac{1}{2}^{+})$ & 37.24 & 116.16 & -41.53 & -72.44 & -39.43 & 1.00 & 0.88 & 1.03 \\\hline
     $1(\frac{3}{2}^{+})$ & 29.63  & 92.50 & -81.32 & -21.67 & -19.15 & 0.91 & 1.36 & 1.40 \\\hline\hline
     \multicolumn{9}{c}{Case III ~~~~~~$\Lambda_{P}=\Lambda_{T}=1.16$ GeV}\\\hline
     $1(\frac{1}{2}^{+})$ & 63.07 & 169.01 & -52.14 & -66.03 & -113.91 & 0.83 & 0.82 & 0.75 \\\hline
     $1(\frac{3}{2}^{+})$ & 46.94 & 141.01 & -61.84 & -25.27 & -100.84 & 0.91 & 1.02 & 0.86\\
    \hline\hline
    \end{tabular}\\
\end{table}

In case I,   the cutoff of the $\bar{D}^{(\ast)}\Sigma_{c}$ potential is taken the same  as that of the $\bar{D}\bar{D}^{\ast}$ potential. For such potentials,   there exist two three-body bound states $\bar{D}\bar{D}^{\ast}\Sigma_{c}$ with $I(J^{P})=1(\frac{1}{2}^{+})$ and $I(J^{P})=1(\frac{3}{2}^{+})$,  and binding energies, 10.9 MeV and 7.1 MeV, respectively. For a  cutoff of $\Lambda=0.998$ GeV, the OBE  $\bar{D}^{\ast}\Sigma_{c}$ potential does not support the existence of $\bar{D}^{\ast}\Sigma_{c}$ bound states corresponding to $P_{c}(4440)$ and $P_{c}(4457)$.   As a result, case I  indicates that there exist two  three-body $\bar{D}\bar{D}^{\ast}\Sigma_{c}$ bound  states even the $\bar{D}^{\ast}\Sigma_{c}$ system does not bind as long as $T_{cc}$ is a $\bar{D}\bar{D}^{\ast}$ bound state.

In case II,  we change the cutoff of the $\bar{D}^{(\ast)}\Sigma_{c}$ potential from 0.998 GeV to 1.16 GeV, while keep the cutoff of the $\bar{D}^{\ast}D^{\ast}$  potential unchanged.  In this case, the strength of the  $\bar{D}^{(\ast)}\Sigma_{c}$  potential becomes stronger, resulting in two  three-body bound states with larger binding energies $37.2$ MeV and $29.6$ MeV. One can see that assuming $T_{cc}$ as a $\bar{D}\bar{D}^{\ast}$ bound state and $P_{c}(4312)$, $P_{c}(4440)$ and $P_{c}(4457)$ as $\bar{D}^{(\ast)}\Sigma_{c}$ bound states, we obtain two three-body bound states below the $\bar{D}\bar{D}^{\ast}\Sigma_{c}$ mass threshold.

\begin{figure}[ttt]
\begin{center}
\includegraphics[width=3.4in]{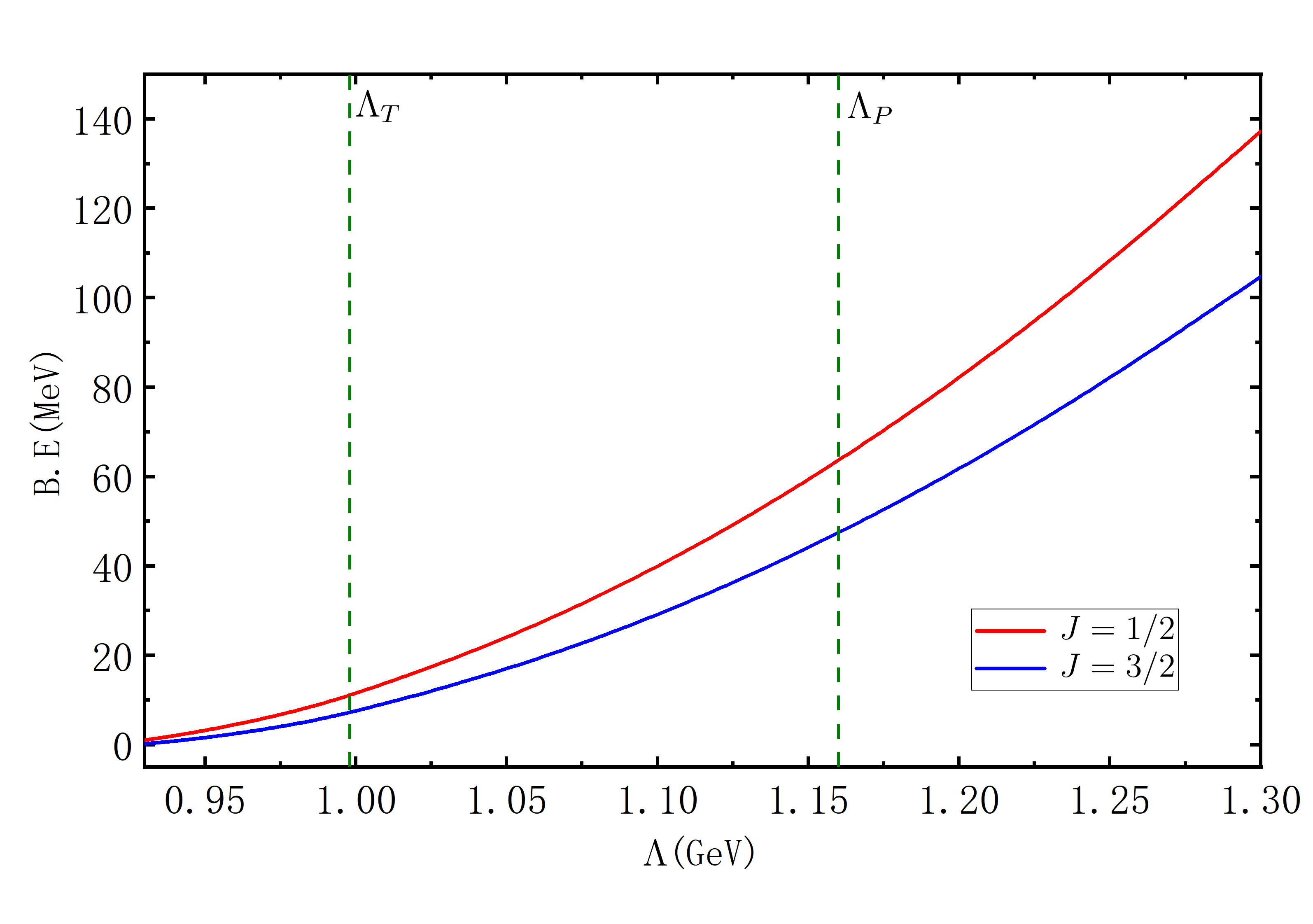}
\caption{Binding energies of the $\bar{D}\bar{D}^{\ast}\Sigma_{c}$ molecules as a function of $\Lambda$. The red and blue solid lines represent the $J^{P}=1/2$ and $J^{P}=3/2$ states, respectively.  The green dot lines represent cutoffs that could  reproduce the binding energy  of $T_{cc}$($\Lambda_{T}$) and the pentaquark states($\Lambda_{P}$).       }
\label{binding energy}
\end{center}
\end{figure}

In case III,  we  change the cutoff of the $\bar{D}^{\ast}\bar{D}$ potential from 0.998 GeV to 1.16 GeV, which naturally results in two  bound states  with even larger binding energies as shown in Table~\ref{results1}. In Fig.~\ref{binding energy}, we present the binding energies  of the $\bar{D}\bar{D}^{\ast}\Sigma_{c}$ system  as a function of $\Lambda$. One can see that the three-body system  $\bar{D}\bar{D}^{\ast}\Sigma_{c}$ remains bound even when both  $DD^{\ast}$ and $\bar{D}^{(\ast)}\Sigma_{c}$ are unbound, with binding energies of the order of  several MeV. If the   $\bar{D}\bar{D}^{\ast}\Sigma_{c}$ bound states are observed experimentally in the future, it will help verify the molecular nature of  $P_{c}(4312)$, $P_{c}(4440)$, $P_{c}(4457)$  and $T_{cc}$ in terms of the results shown in Fig.~\ref{binding energy}.

\begin{figure}[ttt]
\begin{center}
\includegraphics[width=3.5in]{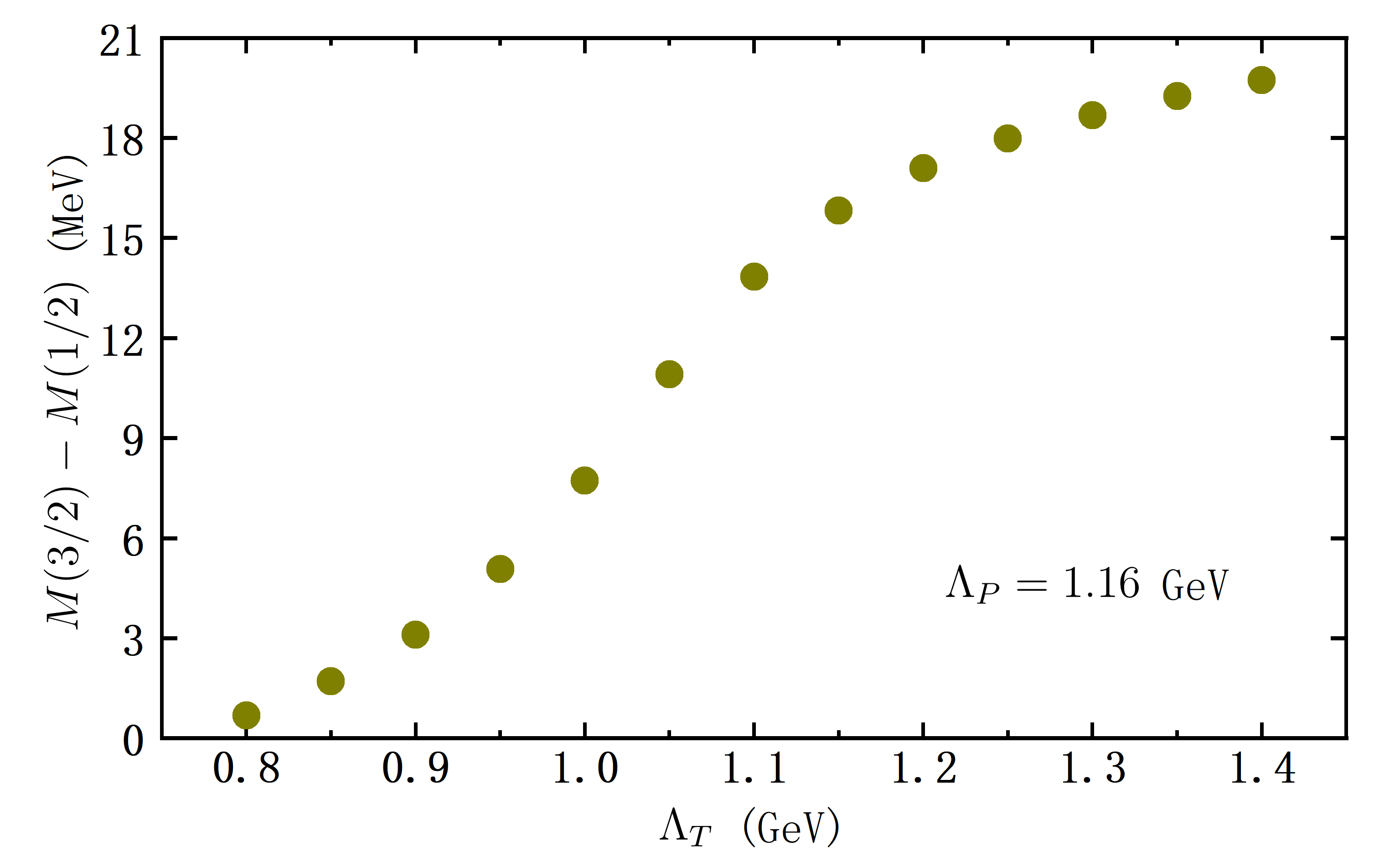}
\caption{ Mass splitting of the three-body $\bar{D}\bar{D}^{\ast}\Sigma_{c}$ doublet  as a function of the cutoff of the $\bar{D}\bar{D}^{\ast}$ potential for fixed $\Lambda_P=1.16$ GeV.   }
\label{mass splitting}
\end{center}
\end{figure}

It is interesting to note that the mass splitting of  the three-body $\bar{D}\bar{D}^{\ast}\Sigma_{c}$ doublet in case III is larger than that of  case II, which implies that the strength of the $\bar{D}\bar{D}^{\ast}$  potential affects the splitting. In Fig.~\ref{mass splitting}, we present the mass splitting as a function of the cutoff of the $\bar{D}\bar{D}^{\ast}$  potential.  It is obvious that the mass splitting increases with the strength of the $\bar{D}\bar{D}^{\ast}$  potential. Interestingly, the mass splitting is 
positive, which means that the mass of the spin 3/2 $\bar{D}\bar{D}^{\ast}\Sigma_{c}$  bound state is  larger than that  of its spin 1/2 counterpart. We note that in this case the $J^{P}=\frac{1}{2}^{-}$ $\bar{D}^{\ast}\Sigma_{c}$ system   is more bound  than the $J^{P}=\frac{3}{2}^{-}$ $\bar{D}^{\ast}\Sigma_{c}$ system. It indicates that the mass splitting of the three-body $\bar{D}\bar{D}^{\ast}\Sigma_{c}$ doublet  is oppositely correlated to the mass splitting of the two-body
$\bar{D}^{\ast}\Sigma_{c}$ bound states, which offers a non-trivial way to check the molecular nature of the involved states. We note in passing that in Ref.~\cite{Pan:2019skd}, we found that the mass splitting of $P_{c}(4440)$ and $P_{c}(4457)$ is correlated to the mass splitting of the $\Xi_{cc}^{(\ast)}\Sigma_{c}^{(\ast)}$ doublet  via heavy antiquark diquark symmetry.

\begin{figure*}[ttt]
\begin{center}
\subfigure[$ ~$]{
\begin{minipage}[t]{0.3\linewidth}
\centering
\includegraphics[width=2.1in]{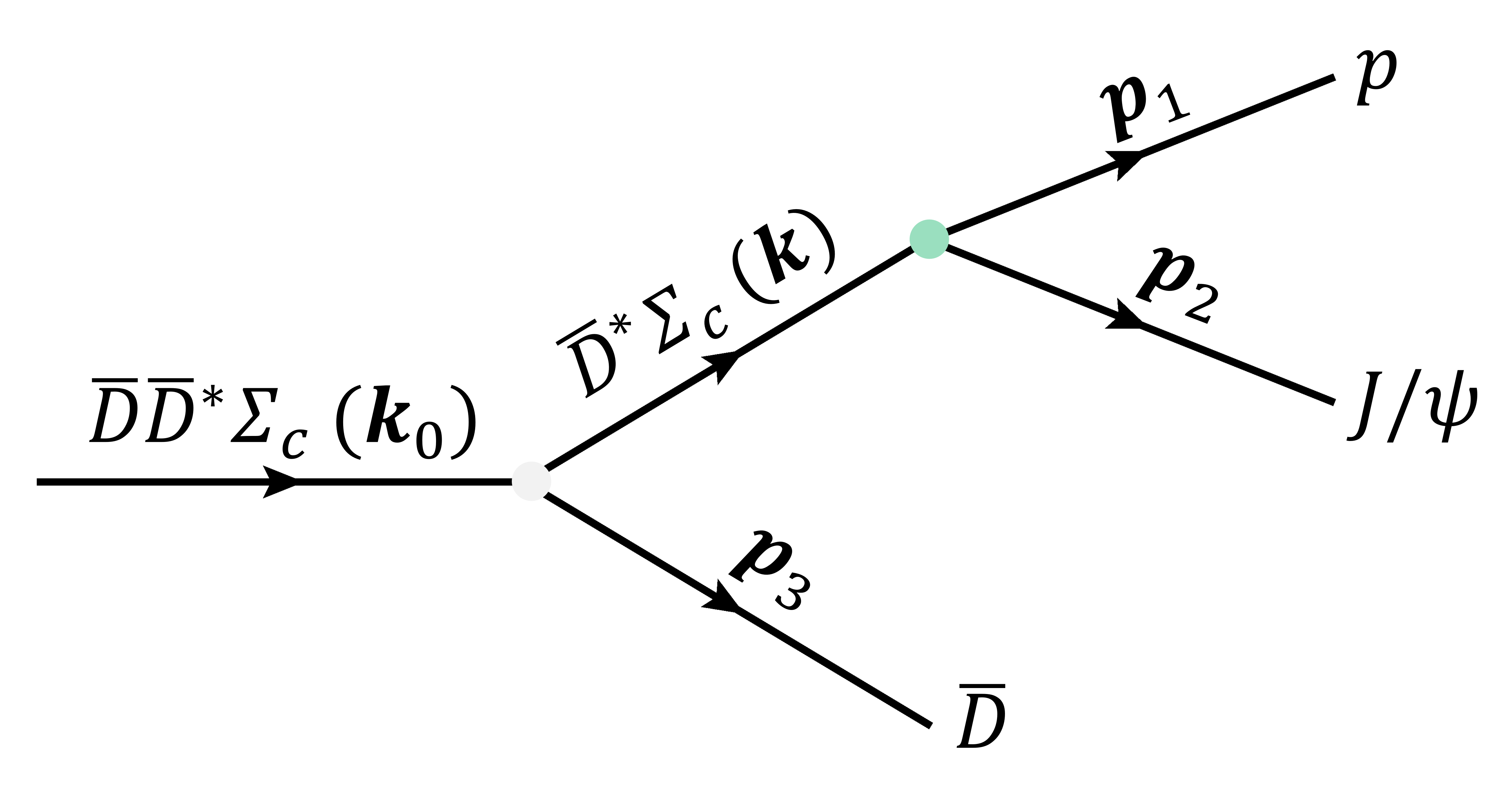}
\end{minipage}
}
\subfigure[$ ~$]{
\begin{minipage}[t]{0.3\linewidth}
\centering
\includegraphics[width=2.1in]{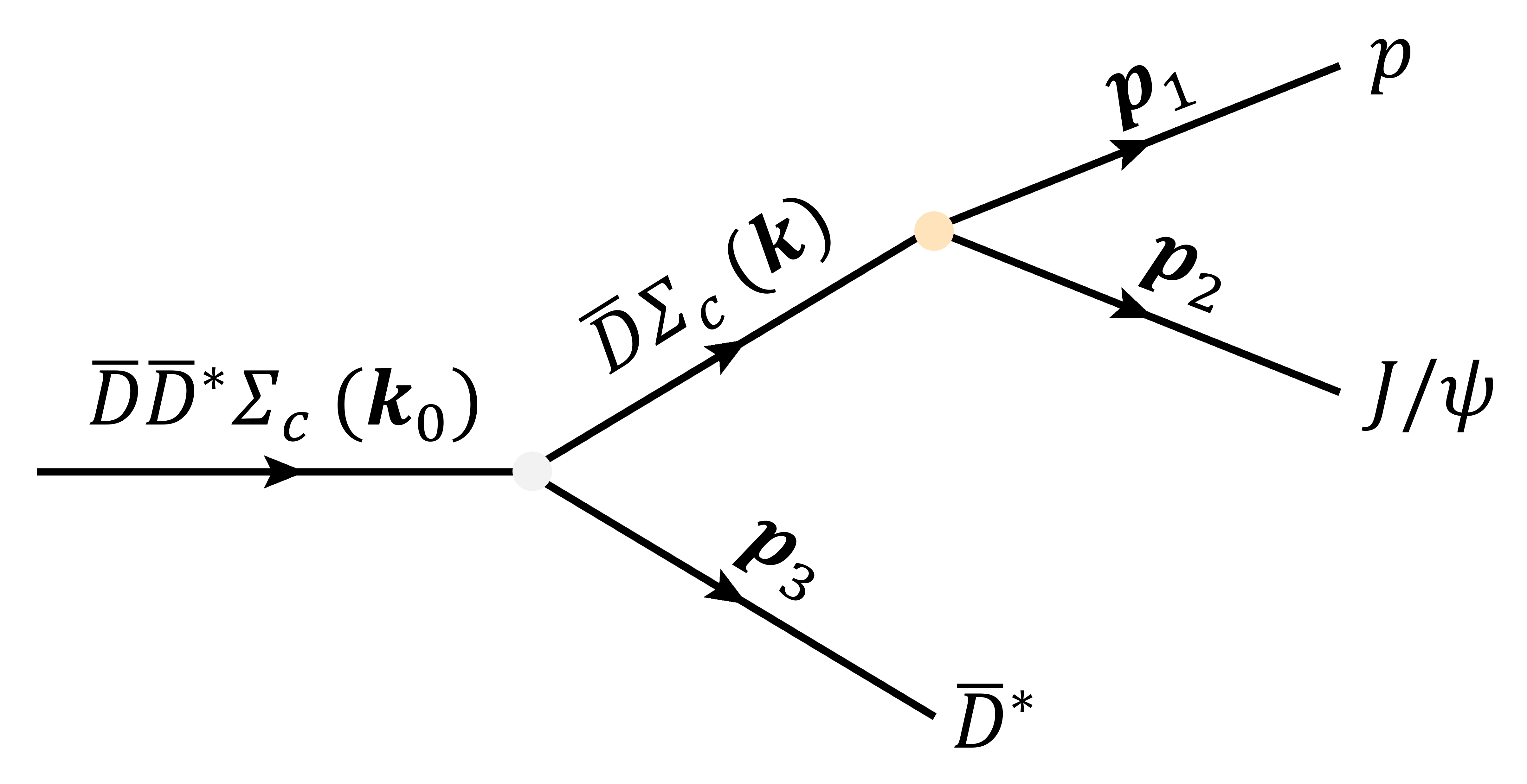}
\end{minipage}
}
\subfigure[$ ~$]{
\begin{minipage}[t]{0.3\linewidth}
\centering
\includegraphics[width=2.1in]{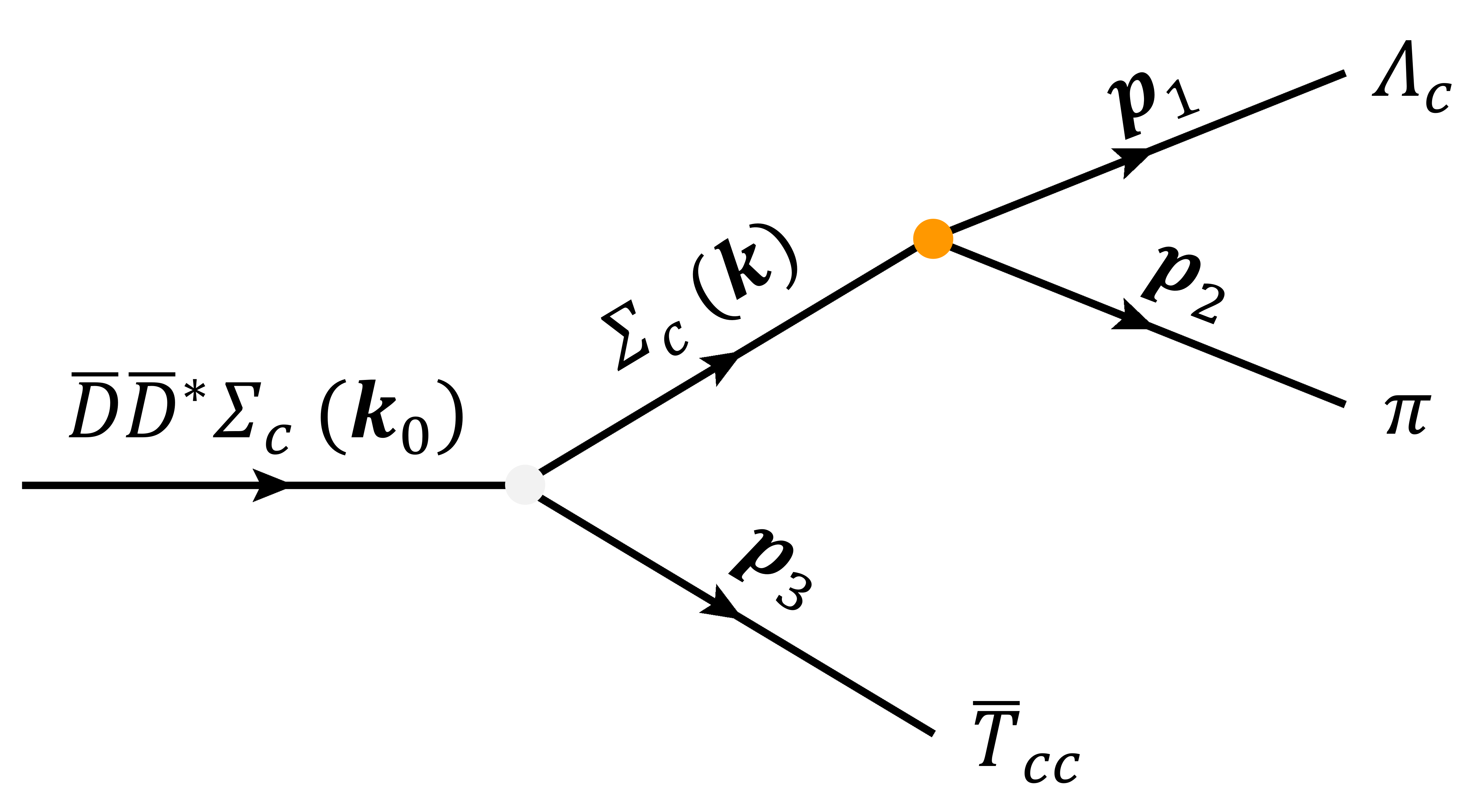}
\end{minipage}
}
\caption{ Tree level diagrams for the three-body   $\bar{D}\bar{D}^{\ast}\Sigma_{c}$ bound states decaying into $J/\psi p \bar{D}$(a), $J/\psi p \bar{D}^{\ast}$(b) and $\pi \Lambda_{c} \bar{T}_{cc}$(c).     \label{decay} }
\label{decay}
\end{center}
\end{figure*}

From our above study we conclude that there exist two three-body  molecules $\bar{D}\bar{D}^{\ast}\Sigma_{c}$ with $I(J^{P})=1(\frac{1}{2}^{+})$ and $I(J^{P})=1(\frac{3}{2}^{+})$. In the following, we denote $P_{c}(4312)$, $P_{c}(4440)$, and $P_{c}(4457)$  as $P_{c1}$, $P_{c2}$, and $P_{c3}$, and the $I(J^{P})=1(\frac{1}{2}^{+})$ and $I(J^{P})=1(\frac{3}{2}^{+})$ $\bar{D}\bar{D}^{\ast}\Sigma_{c}$ bound states as $Hq_{1}$ and $Hq_{2}$, respectively.   We will discuss their possible decay modes and calculate the decay widths via the effective Lagrangian approach.  Such  $\bar{D}\bar{D}^{\ast}\Sigma_{c}$ bound states   can be regarded as three kinds of quasi two-body bound states, $P_{c2}(P_{c3})\bar{D}$, $P_{c1}\bar{D}^{\ast}$ and $\bar{T}_{cc}\Sigma_{c}$.  One should note that the particles, $P_{c2/c3}$, $P_{c1}$, and $\Sigma_{c}$, should be viewed as unstable particles in contrast to the particles,  $\bar{D}$, $\bar{D}^{\ast}$ and $\bar{T}_{cc}$~\cite{ParticleDataGroup:2020ssz,LHCb:2021auc}, and these unstable particles can further decay into  two other particles.   As a result, the three-body bound states
$\bar{D}\bar{D}^{\ast}\Sigma_{c}$ can decay into $J/\psi p \bar{D}$, $J/\psi p \bar{D}^{\ast}$, and $\Lambda_{c}\pi \bar{T}_{cc}$,  
as  shown  in Fig.~\ref{decay}.   Since the minimum number of valence quarks of the $\bar{D}\bar{D}^{\ast}\Sigma_{c}$ states is 7,  it will not couple to a pair of traditional hadrons, which indicates that they can only  decay into   at least three traditional hadrons. In other words, the decay mechanism shown in Fig.~\ref{decay} should be the dominant ones.

In the following, we will show how to calculate the partial decay widths of the $\bar{D}\bar{D}^{\ast}\Sigma_{c}$ bound states in the effective Lagrangian approach.  It should be noted that in the following study, we focus on Case II, among the three cases studied, because it can reproduce the pentaquark states and $T_{cc}$. The effective Lagrangians describing the interactions between the three-body bound states and their constituents have the following form
\begin{widetext}
    \begin{equation}
    \begin{aligned}
     \mathcal{L}_{Hq_{1}P_{c3} \bar{D}}=& g_{Hq_{1}P_{c3} \bar{D}} Hq_{1}(x)   \int dy \bar{D}(x+\omega_{P_{c3}}y) P_{c3}(x+\omega_{\bar{D}}y)\Phi(y^2) ,   \\ 
     \mathcal{L}_{Hq_{2}P_{c2}\bar{D}} =&g_{Hq_{2}P_{c2}\bar{D}} Hq_{2}^{\mu}(x) \int dy  \bar{D}(x+\omega_{P_{c2}}y)P_{c2\mu}(x+\omega_{\bar{D}}y)\Phi(y^2),   \\ 
     \mathcal{L}_{Hq_{1}P_{c1}\bar{D}^{\ast}}=&g_{Hq_{1}P_{c1}\bar{D}^{\ast}} Hq_{1}(x)  \int dy  \bar{D}^{\ast\mu}(x+\omega_{P_{c1}}y)\gamma_{\mu}\gamma_{5}  P_{c1}(x+\omega_{\bar{D}^{\ast}}y)\Phi(y^2),  \\ 
     \mathcal{L}_{Hq_{2}P_{c1}\bar{D}^{\ast}}=&g_{Hq_{2}P_{c1}\bar{D}^{\ast}} Hq_{2\mu}(x) \int dy  \bar{D}^{\ast\mu}(x+\omega_{P_{c1}}y)   P_{c1}(x+\omega_{\bar{D}^{\ast}}y)\Phi(y^2),  \\ 
     \mathcal{L}_{Hq_{1}T_{cc}\Sigma_{c}}=&g_{Hq_{1}T_{cc}\Sigma_{c}} Hq_{1}(x) \int dy  T_{cc}^{\mu}(x+\omega_{\Sigma_{c}}y)\gamma_{\mu}\gamma_{5}  \Sigma_{c}(x+\omega_{\bar{T}_{cc}}y)\Phi(y^2),  \\ 
    \mathcal{L}_{Hq_{2}T_{cc}\Sigma_{c}}=&g_{Hq_{2}T_{cc}\Sigma_{c}} Hq_{2\mu}(x)  \int dy  \bar{T}_{cc}^{\mu}(x+\omega_{\Sigma_{c}}y)\Sigma_{c}(x+\omega_{\bar{T}_{cc}}y)\Phi(y^2), 
    \end{aligned}
    \end{equation}
\end{widetext}
where $\Phi(y^2)$ denotes the Gaussian form factor, $g$ with different subscripts represent the relevant coupling constants, and $\omega_{i}=\frac{m_{i}}{m_{i}+m_{j}}$ is the  kinematical parameter with $m_{i}$ and $m_{j}$ being the masses of the involved hadrons. We rely on the compositeness condition to estimate the above couplings, which is an effective approach to estimate the couplings between bound states and their constituents~\cite{Weinberg:1962hj}.     The condition implies that the coupling constants can be determined from the fact that the renormalization constant of the wave function of a composite particle should be zero. Following our previous works~\cite{Ling:2021lmq,Ling:2021bir}, with the cutoff $\Lambda=1$ GeV and the masses of the three-body bound states  $\bar{D}\bar{D}^{\ast}\Sigma_{c}$  the couplings   are determined as shown in Table~\ref{couplings}.
The details about deriving the couplings can be found in Ref.~\cite{Ling:2021lmq}.

The Lagraingian describing the secondary decay process are expressed as
\begin{eqnarray}
\mathcal{L}_{P_{c1}J/\psi p}&=& g_{P_{c1}J/\psi p} P_{c1} \gamma_{\mu}\gamma_{5} J/\psi^{\mu} p,
\\ \nonumber
\mathcal{L}_{P_{c2}J/\psi p}&=&  g_{P_{c2}J/\psi p} P_{c2\mu}J/\psi^{\mu} p, \\ \nonumber
\mathcal{L}_{P_{c3}J/\psi p}&=&  g_{P_{c3}J/\psi p} P_{c3}\gamma_{\mu}\gamma_{5}J/\psi^{\mu} p,  \\ \nonumber
\mathcal{L}_{\pi\Lambda_{c}{\Sigma}_{c}}&=&\frac{g_{\pi\Lambda_{c}{\Sigma}_{c}}}{f_{\pi}}~\bar{\Lambda}_{c}\gamma^\mu\gamma_5 \partial_\mu\vec{\phi}_\pi\cdot \vec{\tau}{\Sigma}_{c},
\end{eqnarray}
where the pion decay constant $f_{\pi}=132$ MeV and the coupling $g_{\Sigma_{c}\Lambda_{c}\pi}$ are determined as 0.55 by reproducing the decay width of $\Sigma_{c}\to\Lambda_{c}\pi$ of 1.89 MeV~\cite{ParticleDataGroup:2020ssz}.  Since the partial  decay widths of the pentaquark states  into $J/\psi p$ are unknown, we can not determine them in terms of  experimental data. In this work, we determine the couplings of the pentaquark states  to $J/\psi p$  in the contact-range EFT approach, where the $\eta_{c}p$, $J/\psi p$, and $\bar{D}^{(\ast)}\Sigma_{c}$ channels dictated by HQSS are taken into account. By reproducing the masses and widths of the pentaquark states, the relevant couplings are determined to be $g_{p_{c1}J/\psi p }=0.22$, $g_{p_{c2}J/\psi p }=0.44$, and $g_{p_{c3}J/\psi p }=0.33$~\cite{xiejiaming}, consistent with  the chiral unitary approach~\cite{Xiao:2020frg}.

\begin{table}[ttt]
\centering
\caption{Couplings of   the $\bar{D}\bar{D}^{\ast}\Sigma_{c}$ molecules  to their components obtained with a cutoff of $\Lambda=$1 GeV.   }
\label{couplings}
\begin{tabular}{c|ccc}
\hline\hline
Couplings  & $g_{H_{q1} P_{c2} \bar{D}} $  &  $g_{H_{q1}P_{c1}\bar{D}^{\ast}}$  &    $g_{H_{q1}T_{cc}\Sigma_{c}}$   \\
\hline
Value  & 3.04  &  1.70     & 2.58   \\
  \hline\hline
  Couplings  &  $g_{H_{q2}P_{c3}\bar{D}}$&   $g_{H_{q2}P_{c1}\bar{D}^{\ast}}$   &   $g_{H_{q2}T_{cc}\Sigma_{c}} $   \\
\hline
Value   & 1.94 &  2.64  &  4.15  \\
\hline\hline
\end{tabular}
\end{table}


With the above relevant Lagrangians the corresponding amplitudes of the strong  decays of  Fig.~\ref{decay} are  
\begin{widetext}
    \begin{equation}
    \begin{aligned}
     \mathcal{M}_{a(J=1/2)}=& i g_{Hq_{1}\bar{D} P_{c3}} g_{P_{c3}J/\psi p}\bar{u}_{Hq_{1}}\frac{1}{{/\!\!\! k}-m_{P_{c3}}}\gamma_{\mu}\gamma_{5}\varepsilon^{\mu}(p_{2}) u_{p}, \\  \mathcal{M}_{a(J=3/2)}=& i g_{Hq_{2}\bar{D} P_{c2}} g_{P_{c2}J/\psi p}\bar{u}_{Hq_{2}}^{\mu}\frac{S_{\mu\nu}(k)}{{/\!\!\! k}-m_{P_{c2}}}\varepsilon^{\mu}(p_{2}) u_{p},  \\ \mathcal{M}_{b(J=1/2)}=& i g_{Hq_{1}\bar{D}^{\ast} P_{c1}} g_{P_{c1}J/\psi p}\bar{u}_{Hq_{1}}\varepsilon_{\nu}(p_{3})\gamma^{\nu}\gamma^{5}  \frac{1}{{/\!\!\! k}-m_{P_{c1}}}\gamma_{\mu}\gamma_{5}\varepsilon^{\mu}(p_{2}) u_{p},   \\ \mathcal{M}_{b(J=3/2)}=& i g_{Hq_{2}\bar{D}^{\ast} P_{c1}} g_{P_{c1}J/\psi p}\bar{u}_{Hq_{1}}^{\nu}\varepsilon_{\nu}(p_{3})\frac{1}{{/\!\!\! k}-m_{P_{c1}}}  \gamma_{\mu}\gamma_{5}\varepsilon^{\mu}(p_{2}) u_{p},  \\ 
     \mathcal{M}_{c(J=1/2)}=& i g_{Hq_{1}T_{cc}\Sigma_{c}}  \frac{g_{\pi\Lambda_{c}{\Sigma}_{c}}}{f_{\pi}}\bar{u}_{Hq_{1}}\varepsilon_{\nu}(p_{3})\gamma^{\nu}\gamma^{5}\frac{1}{{/\!\!\! k}-m_{\Sigma_{c}}}  \gamma_{\mu}\gamma_{5}p_{2}^{\mu} u_{\Lambda_{c}},
\\ 
\mathcal{M}_{c(J=3/2)}=& i g_{Hq_{2}T_{cc}\Sigma_{c}}  \frac{g_{\pi\Lambda_{c}{\Sigma}_{c}}}{f_{\pi}}\bar{u}_{Hq_{2}}^{\nu}\varepsilon_{\nu}(p_{3})\frac{1}{{/\!\!\! k}-m_{\Sigma_{c}}}  \gamma_{\mu}\gamma_{5}p_{2}^{\mu} u_{\Lambda_{c}},
    \end{aligned}
    \end{equation}
\end{widetext}
where $u$ and $\bar{u}$ represent the corresponding spinor function denoted by the subscripts, $\varepsilon$ is the polarization vector, and $S_{\mu\nu}=g^{\mu\nu}-\frac{1}{3}\gamma^{\mu}\gamma^{\nu}-\frac{\gamma^{\mu}p^{\nu}-\gamma^{\nu}p^{\mu}}{3m}-\frac{2p^{\mu}p^{\nu}}{3m^2}$ with $m$ being the corresponding mass of an spin-3/2 particle.

With the so-obtained amplitudes , one can easily  calculate the partial decay width
\begin{equation}
d\Gamma =  \frac{1}{(2 \pi)^{3}}\frac{1}{2J+1} \frac{\overline{|\mathcal{M}|^2}}{32 m_{Hq_{1(2)}}^{3}} d m_{12}^{2} d m_{23}^{2}.
\end{equation}

In Table~\ref{results}, we present the partial decay widths of the two three-body bound states, $Hq_{1}$ and $Hq_{2}$. We find that the bound state with $J=1/2$ dominantly decays into $J/\psi p \bar{D}$,  much larger than that of the  $J=3/2$ state. Therefore, the $J/\psi p\bar{D}$ mode is a golden channel to discriminate the spin of $Hq_{1}$ and $Hq_{2}$ molecules. $Hq_{1}$ and $Hq_{2}$   decay almost equally into $J/\psi p \bar{D}^{\ast}$, while  the decay into $\bar{T}_{cc}\Lambda_{c}\pi$ is rather small in contrast to the other two decay modes. One should note that  the partial decay widths of the three pentaquark states decaying into $J/\psi p$ are not  very precisely known~\cite{Xiao:2020frg}, leading to some uncertainties about the partial decay widths given in Table~\ref{results}. Nonetheless, we suggest  to search for them in the $J/\psi p \bar{D}^{\ast}$ or $J/\psi p\bar{D}$ mass distributions.        

\begin{table}[!h]
\centering
\caption{Partial decay widths of the $\bar{D}\bar{D}^{\ast}\Sigma_{c}$ molecules.   }
\label{results}
\begin{tabular}{c|cccccccccc}
\hline\hline
Modes of  Fig.~\ref{decay}(a)  & $Hq_{1} \to J/\psi p\bar{D}  $  &   $Hq_{2}  \to J/\psi p  \bar{D}$    \\
Value (MeV)  & 12.3  &  0.9       \\ \hline 
Modes of  Fig.~\ref{decay}(b)   &  $Hq_{1} \to J/\psi p \bar{D}^{\ast} $  &  $Hq_{2}  \to  J/\psi p \bar{D}^{\ast}$   \\    
Value (MeV)  & 3.7  & 3.9   \\
\hline
Modes  of  Fig.~\ref{decay}(c)  &  $Hq_{1}  \to \bar{T}_{cc}\Lambda_{c}\pi$     &  $Hq_{2} \to \bar{T}_{cc}\Lambda_{c}\pi  $     \\ 
Value (keV) & 0.2 & 3.3     \\     
  \hline\hline
\end{tabular}
\end{table}

\section{Summary and conclusion}
\label{sum}

The exotic states, $T_{cc}$, $P_{c}(4312)$, $P_{c}(4440)$, and $P_{c}(4457)$,  discovered by the LHCb Collaboration recently, have been suggested to be  $DD^{\ast}$
and $\bar{D}^{(\ast)}\Sigma_{c}$ hadronic molecules. However, their molecular nature is difficult to be confirmed either experimentally or theoretically. In this work, we have investigated these exotic states in the three-body $\bar{D}\bar{D}^{\ast}\Sigma_{c}$ system, which is equivalent to  adding a $\bar{D}$ meson into the  $\bar{D}^{\ast}\Sigma_{c}$ system. The OBE interactions of the sub-systems, $\bar{D}^{(\ast)}\Sigma_{c}$ and $\bar{D}\bar{D}^{\ast}$,  are determined by reproducing the masses of the molecular candidates, i.e., the three pentaquatk states [$P_{c}(4312)$, $P_{c}(4440)$ and $P_{c}(4457)$] and $\bar{T}_{cc}$.   After solving the three-body schr\"odinger equation, we obtained two three-body  bound states,  $I(J^{P})=1(\frac{1}{2}^{+})$ $\bar{D}\bar{D}^{\ast}\Sigma_{c}$ and $I(J^{P})=1(\frac{3}{2}^{+})$ $\bar{D}\bar{D}^{\ast}\Sigma_{c}$, with binding energies 37.2 MeV and 29.6 MeV, respectively. In particular, we explored the correlation between the existence of $\bar{D}\bar{D}^{\ast}\Sigma_{c}$ molecules  with the existence of $\bar{D}^{(\ast)}\Sigma_{c}$ and $\bar{D}^{\ast}\bar{D}$ molecules.  If the $\bar{D}\bar{D}^{\ast}\Sigma_{c}$ bound states  can be observed experimentally in the future,  the correlation can help to test the molecular nature of $P_{c}(4312)$, $P_{c}(4440)$,  $P_{c}(4457)$ and $\bar{T}_{cc}$.
The mass splitting of the three-body doublet is found to be correlated to that of the $\bar{D}^{\ast}\Sigma_{c}$ doublet. Assuming  $P_{c}(4457)$ and $P_{c}(4440)$ as $J=1/2$ and $J=3/2$ $\bar{D}^{\ast}\Sigma_{c}$ bound states, respectively, we find that the  mass splitting between $I(J^{P})=1(\frac{1}{2}^{+})$ and  $I(J^{P})=1(\frac{3}{2}^{+})$ $\bar{D}\bar{D}^{\ast}\Sigma_{c}$ bound states is positive. 

At last, we employed the effective Lagrangian approach to calculate the partial  decay widths of $\bar{D}\bar{D}^{\ast}\Sigma_{c}$ bound states. We find that $J=1/2$ and $J=3/2$ $\bar{D}\bar{D}^{\ast}\Sigma_{c}$ bound states mainly decay into $J/\psi p \bar{D}$ and $J/\psi p \bar{D}^{\ast}$, respectively, while the decay into $\bar{T}_{cc}\Lambda_{c}\pi$ is small. We strongly recommend experimental  searches for such three-body bound states in the $J/\psi p \bar{D}$ and $J/\psi p \bar{D}^{\ast}$  mass distributions, which can help verify the molecular nature of  $T_{cc}$, $P_{c}(4312)$, $P_{c}(4440)$, and $P_{c}(4457)$.

\section{Acknowledgments}
  This work is supported in part by the National Natural Science Foundation of China under Grants No.11975041,  No.11735003, and No.11961141004. Ming-Zhu Liu acknowledges support from the National Natural Science Foundation of
China under Grant No.1210050997.  Tian-Wei Wu acknowledges support from the National Natural Science Foundation of China under Grant No.12147152.

\bibliography{dds}

\end{document}